\documentclass[preprint]{revtex4}

\usepackage{graphicx,psfrag,xcolor,bm}

\begin{document}

\title{Portfolio Construction Matters}

\author{Stefano Ciliberti and Stanislao Gualdi}
\affiliation{Capital Fund Management \\
  23 rue de l'Universit\'e, 75007 Paris, France}

\date{\today}

\begin{abstract}
  The role of portfolio construction in the implementation of equity
  market neutral factors is often underestimated. Taking the classical
  momentum strategy as an example, we show that one can significantly
  improve the main strategy's features by properly taking care of this
  key step. More precisely, an optimized portfolio construction
  algorithm allows one to significantly improve the Sharpe Ratio,
  reduce sector exposures and volatility fluctuations, and mitigate
  the strategy's skewness and tail correlation with the market. These
  results are supported by long-term, world-wide simulations and will
  be shown to be universal. Our findings are quite general and hold
  true for a number of other ``equity factors''. Finally, we discuss
  the details of a more realistic set-up where we also deal with
  transaction costs.
\end{abstract}

\maketitle


\section{Introduction}\label{intro}

A portfolio manager in charge of the implementation of a systematic
equity strategy faces a number of challenges ranging from alpha signal
research, allocation, risk and cost management, and portfolio
construction. In this paper, we assume that alpha signals are available
(taken for instance from the large literature on equity factors), and
that a certain weighting scheme has been chosen. We then study the
role of portfolio construction: more precisely, we are interested in
how the information contained in a factor (which we will equivalently refer
to as a predictor or signal) is translated
into physical trades and positions.

As is well known, this problem was first tackled in a quantitative way by
Markowitz~\cite{Marko_52} in 1952 in what is considered to be the first
brick of Modern Portfolio Theory (MPT)~\footnote{A similar approach was independently
proposed by Roy~\cite{Roy1952}}. From a theoretical
perspective, that paper set the scene for the research work of
Sharpe~\cite{sharpe65}, Treynor~\cite{treynor1965rate}, Ross~\cite{ross1976}, Roll~\cite{fama1969adjustment}, 
Fama and French~\cite{fama1965,malkiel1970efficient}, Chen~\cite{chen1983exact}
and many others~\cite{bla92,merton1969lifetime,Merton1973}, that led to the development of the Capital Asset Pricing Model and later on
the Arbitrage Pricing Theory~\cite{ross1976,roll1980empirical} and factor
modeling~\cite{Fama_French_93,Carhart1997}.

For the financial practitioner, the interesting statistical observation put forward by Markowitz is that
by combining assets whose correlation is less than 1, one can
significantly reduce the risk of the total portfolio. In more
mathematical terms, the problem of maximising the expected gain under
a risk budget constraint translates into a mean-variance portfolio
construction algorithm. When applied to real-world situations, though,
the Markowitz solution often ends up loading on a relatively small
number of (supposedly) weakly correlated stocks. In practice, the
mathematical solution is a function of the inverse of the stocks'
covariance matrix which is known to be very noisy, that is, strongly
dependent on the measurement protocol. Although a number of techniques
has been proposed to ``clean'' the covariance matrix and shown to be
quite efficient (see e.g.~\cite{BUN20171} for a recent review), the lack of
robustness of the naive Markowitz solution is probably the main reason
why the equal-weighted ``$1/N$'' approach has gained more popularity in the
subsequent decades. As the story goes, Harry Markowitz himself would
not apply the efficient frontier criterion to his personal portfolio,
instead going for the simplest $1/N$ rule~\cite{zweig2007your}.

Since then, academic research has been more devoted to the
definition and identification of equity factors than to portfolio
construction. Broadly speaking, equity factors can be classified into
three different categories: 
\begin{itemize}
\item Macroeconomic (exogenous) factors like oil
prices, industrial production, inflation or surprises in the yield
curve~\cite{ChenRollRoss86}; 
\item Statistical factors, which can be extracted
from empirical covariance matrices through statistical techniques such
as the Principal Component Analysis (PCA); and 
\item Microeconomic or Fundamental (endogenous) factors, built as long-short equity
portfolios aimed at capturing features like fundamental ratios,
price-based indicators, or even industry or country membership. 
\end{itemize}
The latter factors are by far the most popular and we will concentrate exclusively
on them. They include the well-known Value, Size, Momentum,
Low-Volatility and Quality factors, among others (see e.g.~\cite{Harvey_al_15,schwert2003anomalies}).

In the fist part of this paper we will pay particular attention to the
Momentum factor~\cite{Roy1952,JegadeeshTitman,MoskowitzOoi2012,DanielMoskowitzMomentum}. 
In a nutshell, a Momentum strategy consists
in buying past winners and selling past losers. It is one of the
oldest and most popular strategies in the financial market, and it has
been proved to be successful across different asset classes,
geographies, and time periods. The equity market-neutral Momentum
strategy is implemented in the single-name equity space and it
corresponds to a long-short relative value portfolio which takes no
net exposure to the market as a whole. The signal is usually defined
as the cumulative performance of a stock over the past 12 months,
where the last month is taken away~\cite{Carhart1997}. Put differently,
the momentum signal needs to be lagged by (at least) one month in
order to get rid of the mean-reversion effect which is quite strong on
that time-scale and as such would deteriorate the trend following performance.  
As our focus is on the role of the portfolio construction, the precise
definition of the signal is not important in our study and we will
stick to the classical academic definition mentioned above.

Our main results can be shortly summarized as follows: In order to get
statistically significant results, robustness across regions and
epochs, and better performance profiles, the portfolio construction
step plays a crucial role. The Markowitz solution, when the
correlation cleaning is correctly taken care of, leads to better
results for the main equity factors in almost all cases we considered. 


\section{Setting the stage: Data, Signal, Risk, and Methodology} \label{setup}

We will use daily stock data in the period 1996-2018 (sources:
Bloomberg and Reuters) on the following geographical zones:
\begin{itemize}
\item US: the $1000$ most liquid stocks in the Russell 3000
  Index. More in detail, we compute the aggregated Average Daily
  (dollar) Volume over $3$ months and take the $1000$ most liquid
  stocks as our investment pool for the next quarter.  This puts us in
  a realistic, causal set-up. The same liquidity-based procedure is
  applied to the below pools.
\item Canada: we first collect quarterly data about company
  fundamentals from Bloomberg and select the $500$ largest cap
  stocks from which we extract the $200$ most liquid stocks.
\item Europe: this zone includes the UK in addition to developed markets in
  continental Europe. Here too we first take the $2000$ largest cap stocks and 
then select the $1000$ most liquid stocks among them.
\item Japan: the $500$ most liquid stocks in the TOPIX index.
\item Australia: we take the $500$ largest cap stocks and then
  select the $200$ most liquid stocks among them.
\end{itemize}

We also independently back-test our strategies on CRSP data, that is US daily
data since 1927, in order to get a higher level of statistical
significance. Here as well, we build a pool with the 1000 most liquid
stocks (to note, there are less than 1000 stocks in the CRSP data base
until about 1950, but this number is always larger than 1000 since
then). The annualized volatility of the performance will be
chosen as the risk metrics, but special attention will be devoted to
tail events and tail correlation with the market.  We will assume to
have a normalized, ranked predictor $p_i$ on stock $i$ that is uniformly distributed in
$[-1,+1]$ and will test the following (known) techniques.

\begin{itemize}
\item[FF:] Following Fama and French (FF), we put on the
  buy-list the top $30\%$ of the stocks, on the sell-list the bottom
  $30\%$, and take long or short positions accordingly. These
  positions are taken proportionally to market-cap.
\item[Neutral:] We take a position $x_i$ on all the stocks $i$,
  proportionally to the predictor: $x_i \propto p_i$. The resulting
  portfolio is cash neutral by construction.  This portfolio is
  supposed to be more diversified than FF, and positions evolve more continuously
  with time.
\item[Beta:] In the two previous examples, market neutrality is
  not necessarily ensured because the single-stocks' sensitivities to
  the market (i.e. the $\beta$'s) may be different. One simple way to
  deal with that is to compute the aggregate $\beta$ of the long and
  short legs separately, and rescale them such that the overall
  $\beta$ is zero. Here the $\beta$ is computed from a simple 1-year
  statistical regression of a one-factor model, where the driving
  factor is identified with the market-cap weighted index. 
\item[Betaopt:] Another way to implement the $\beta$ neutrality while
  maximising the expected gain as defined by the predictor is to solve
  the following optimization problem:
\begin{equation}
  \min_{\{\bm{x}\}}\quad (\bm{p}-\bm{x})^T C (\bm{p}-\bm{x})\quad\quad
  \mbox{subject to}\quad \bm{\beta}\cdot\bm{x} = 0
  \label{Betaoptim}
\end{equation}
where $C_{ij}$ is the covariance matrix of the stocks' returns and
everything is in vectorial notation such that $(\bm{x})_i=x_i$. The
rationale behind the optimization problem in (\ref{Betaoptim}) is that
we are looking for a portfolio which is as close as possible to $x_i =
p_i$ while respecting the Beta neutrality. The solution to the above
problem is given by
\begin{equation}
  \bm{x} = \bm{p} - \frac{\bm{\beta}\cdot \bm{p}}{\bm{\beta}^T C^{-1}\bm{\beta}}C^{-1}\bm{\beta}
\end{equation}
which, by substituting $\beta\propto C\bm{w}$ (where $\bm{w}$ is the
weight vector in the index), can be conveniently rewritten as
\begin{equation}
  \bm{x} = \bm{p} - \frac{\bm{w}^TC\bm{p}}{\bm{w}^TC\bm{w}}\bm{w}
\end{equation}
where no matrix inversion is needed~\footnote{For ``fairness'' we however use the
same cleaned covariance matrix as for the Markowtz construction.}.
Note that if $\bm{\beta}$ coincides with the first eigenvector of the
covariance matrix $C$ then the portfolio is simply built as a
predictor whose projection to the $\bm{\beta}$ vector is removed. 
 
\item[Markowitz:] The classical solution to the portfolio construction
  problem is (at least formally) given by the Markowitz solution, that
  is $x_i \propto C^{-1} p_i$. As already mentioned in the
  introduction, the empirical covariance matrix is extremely noisy and
  has to be ``cleaned'' before the inversion. Here, we only keep the
  $k$ largest eigenvectors of the covariance matrix, which correspond
  to the so-called statistical factors.
\end{itemize}
The covariance matrix is computed empirically from individual stock
returns, usually taking at least two years of historical
close-to-close daily returns (see e.g. \cite{BUN20171} for a discussion on
how long the time series should be for a pool of stock of a certain
size). That matrix is then diagonalized and the top eigenvectors
(that is, the eigenvectors corresponding to the top eigenvalues) are
extracted: these are called ``modes'' or ``statistical factors''. 
It is interesting to elaborate more on the role of the statistical
factors in the development of the Markowitz solution. The $k=1$
factor, also called the ``market mode'', can be seen as a portfolio
where most if not all stocks are aligned in the same direction (more
or less so, depending on their sensitivity to a macroeconomic move,
that is, their individual $\beta$'s). As a consequence, the
corresponding Markowitz solution is close to the $\beta$-neutral cases
above. The factors for $k>1$ are less universal and are typically
related to industry sectors, or to groups driving the local
economy~\cite{Plerou,UechiStanley}. For example, the $k=2$ statistical factor of an Australian
pool of stocks usually exhibits a large (say) positive exposure to
Basic Materials and/or Energy stocks, and a negative exposure to
Financial stocks. The $k=2$ statistical factor on another pool is in
general different (it could be UK vs continental stocks in Europe, for
instance, or Consumer-Non-Cyclical vs Consumer-Cyclical in Japan). The
idea behind the eigenvalue methodology is that one can let the
covariance data speak and identify the main risk drivers on any pool
of stocks, which themselves may change in time. The Markowitz solution
will then automatically reduce the portfolio's exposure to these risk
drivers accordingly.  The choice of $k$ can be optimized in different
ways~\cite{BUN20171}, but we will instead test a few values up to $k=5$ in
order to better understand its role in the present study.

In more formal terms, we will do the following approximation: 
\begin{equation}
  C_{ij} \simeq \sigma_i \sigma_j \bigg(\sum_{\alpha=1}^k \lambda^{\alpha} v^{\alpha}_i
  v^{\alpha}_j  + \varepsilon^2 \delta_{ij}\bigg)
  \label{clip}
\end{equation}
where $\sigma_i$ is the volatility of stock $i$, $\lambda^{\alpha}$,
and $\bm{v^{\alpha}}$ are respectively eigenvalues and eigenvectors
of the correlation matrix, and the diagonal variance term
$\varepsilon^2$ is chosen such that the total variance of the
portfolio is unchanged, i.e. Tr$C = \sum_i \sigma_i^2$.  

In any of the above setups, the portfolio is rebalanced on a daily
basis. The world-wide portfolio is built as a flat average
of the various local portfolios~\footnote{Since there are no costs
involved we prefer a flat average but all the results of the paper
are unaffected if we use a market-cap weighted average of different geographical
zones.}. 
In order to compare the different
portfolio construction methods, each world-wide portfolio has been
rescaled in order to have the same average risk over the total
back-test period.

\begin{figure}[]
   \includegraphics[width= 0.45\textwidth]{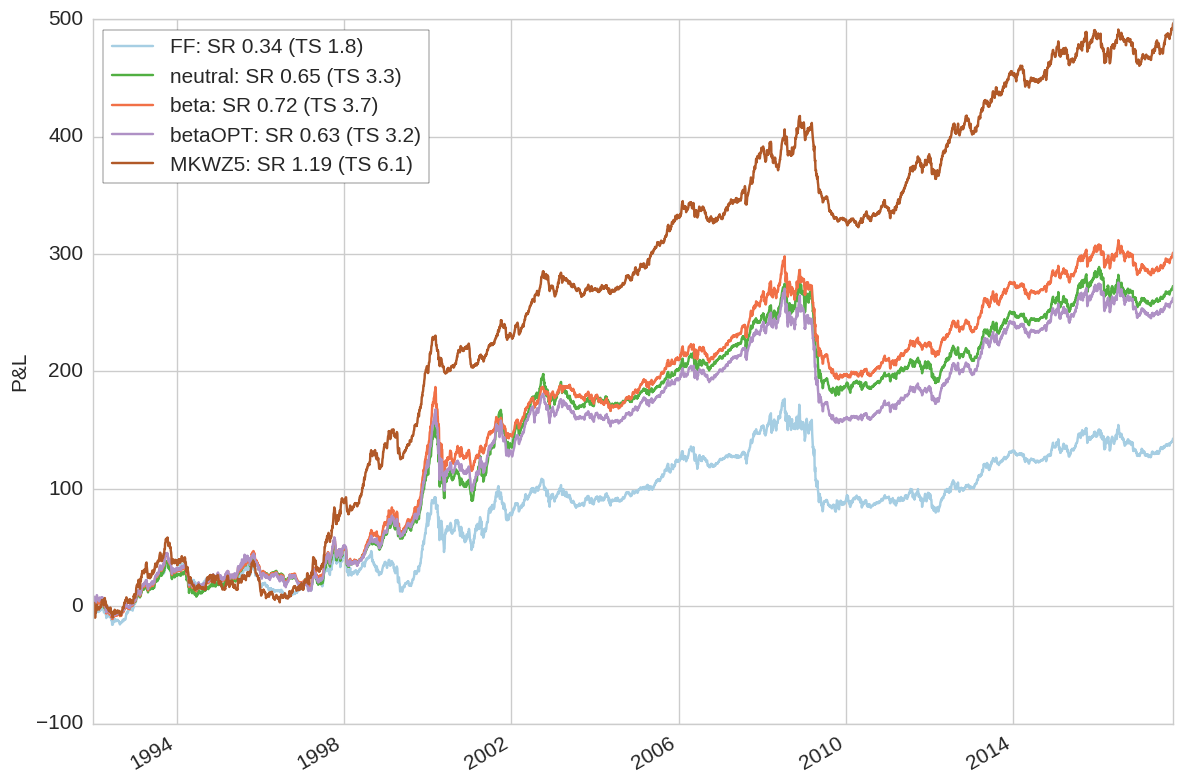}
   \includegraphics[width= 0.45\textwidth]{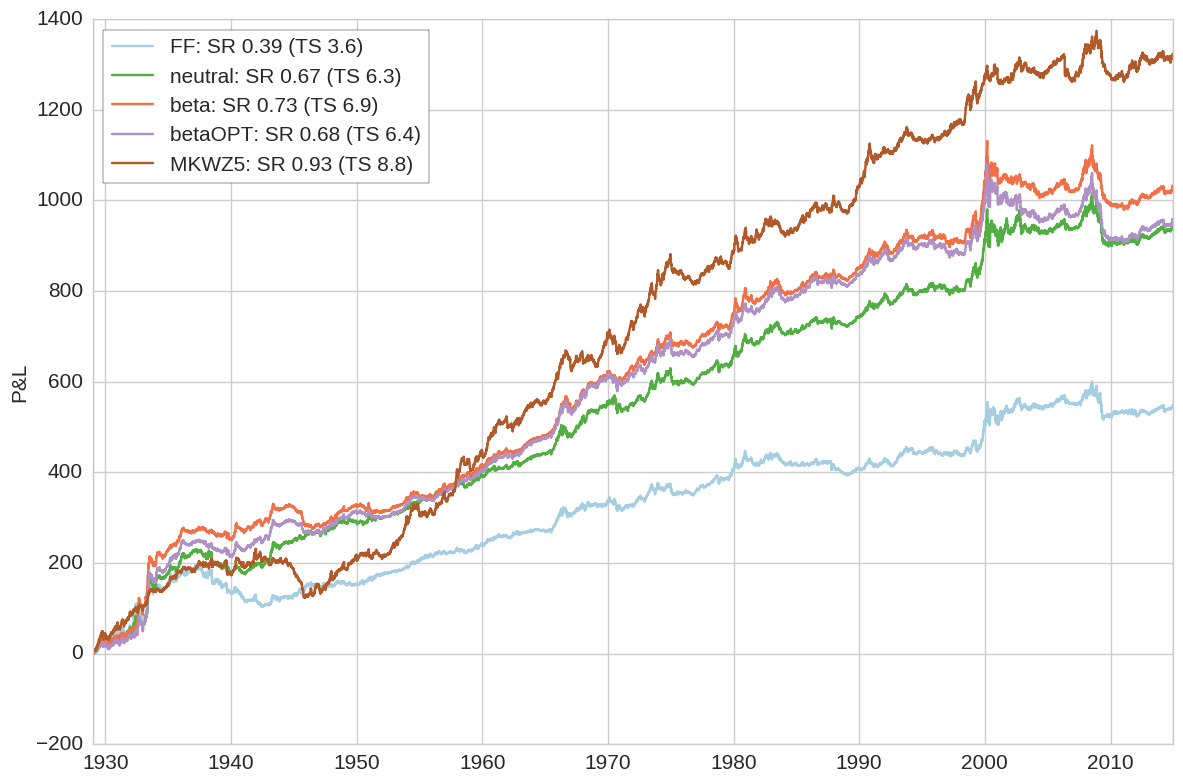}
   \caption{Left: The world-wide performance of the momentum strategy
     with different portfolio construction techniques. Right: Same as
     in the left panel, but on the CRSP US pool since 1927. }
   \label{PnLWW}
\end{figure}

\section{Results for the Momentum Strategy}

In this section we analyze the results obtained for the Momentum
strategy. The left panel of Fig.~\ref{PnLWW} shows the world-wide P\&L 
corresponding to the different portfolio construction algorithms
described in the previous section. The FF construction is the worst
performing with a Sharpe Ratio of $0.34$, while the Markowitz solution,
with $k=5$ statistical factors of the covariance matrix kept before matrix inversion, 
has the highest Sharpe Ratio ($1.19$). All the
other schemes are in between these two extremes. In term of
statistical significance, the t-stat of the P\&L  of the Markowitz
solution is well above the usual acceptance threshold (i.e. t-stat
larger than $3$). The cash-neutral and $\beta$-neutral solutions are
borderline significant, while the FF is below (t-stat=$1.8$). These
summary numbers are very close to those found on a much longer
back-test ranging from 1927 until 2016 (right-panel) where we use CRSP
data for US stocks.

\begin{figure}[]
   \includegraphics[width= 0.45\textwidth]{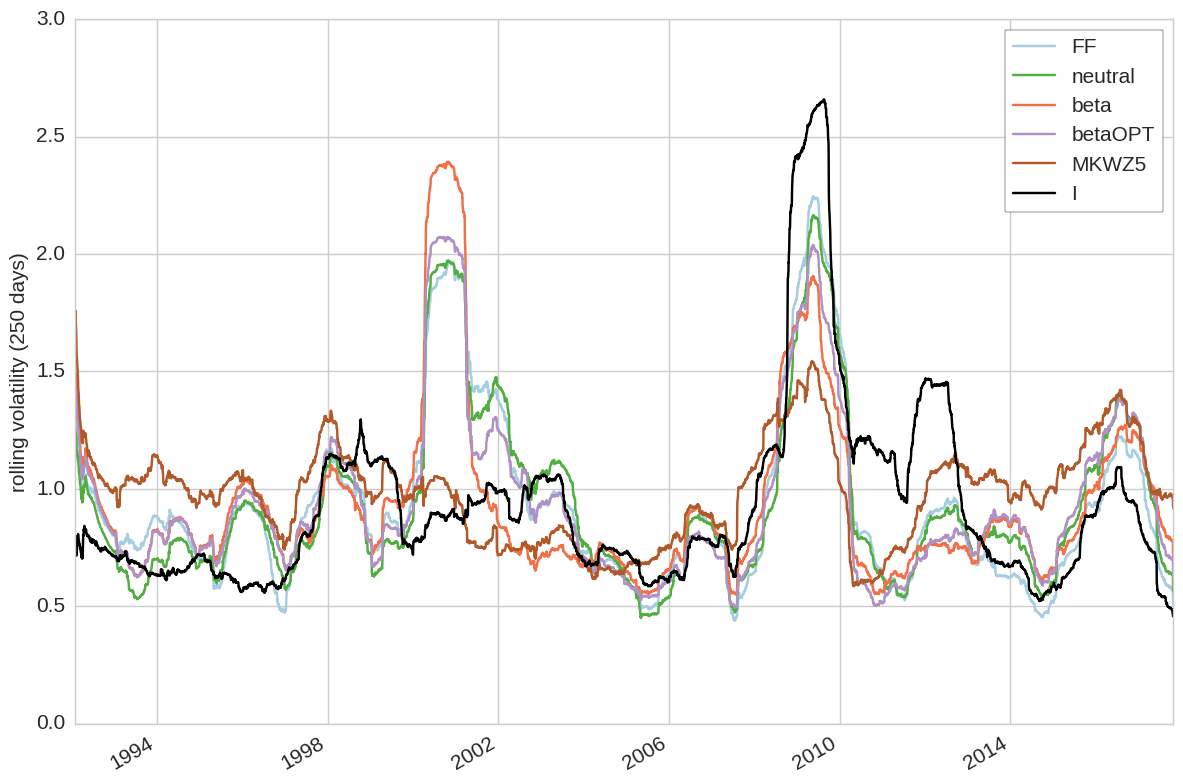}
   \includegraphics[width= 0.45\textwidth]{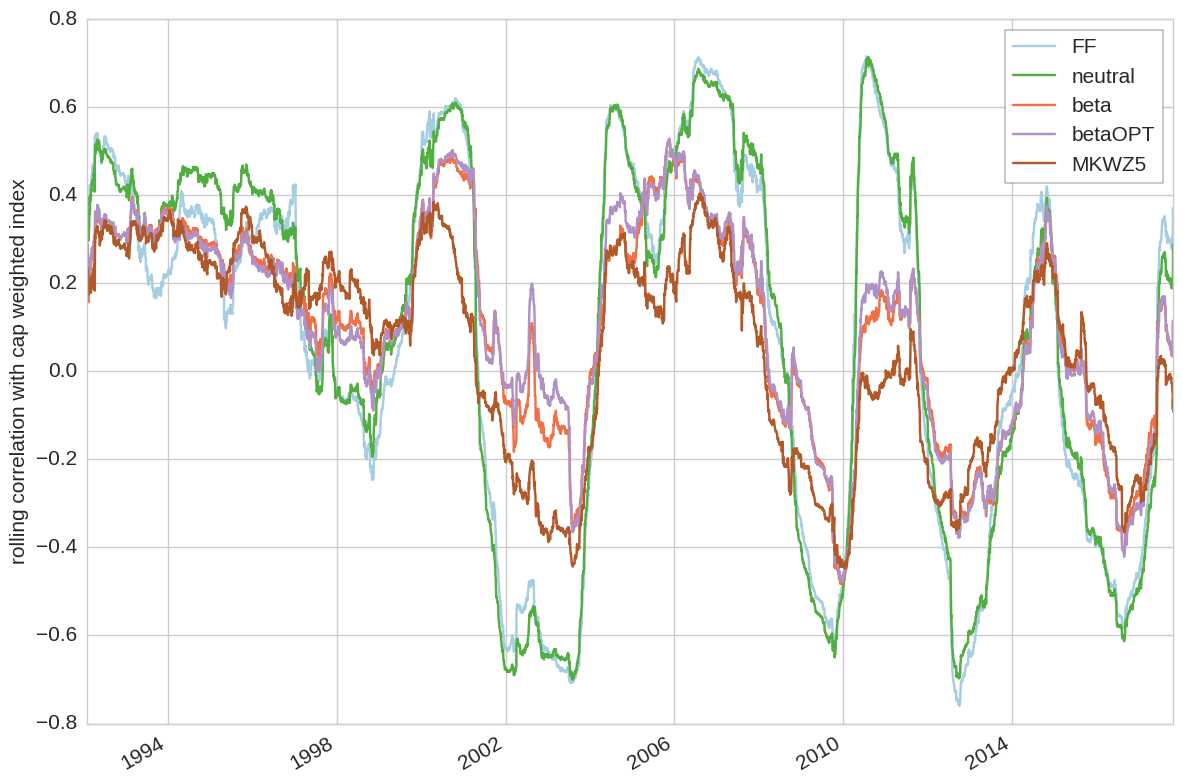}
   \caption{Left panel: The 1-year rolling volatility of the different
     P\&Ls in Fig.~\ref{PnLWW}. The time-average is set to 1 for all
     and fluctuations are within a factor 2.  Right panel: The 1-year
     rolling correlation between the performance and the market-cap
     weighted index. The time-average of these curves is always
     consistent with zero, but fluctuations can be pretty large over
     certain time periods. }
   \label{rollvol}
\end{figure}

In order to get a better understanding of why the Markowitz solution
works better than classical score-based approaches, we look into the
details of the risk profile. The 1-year rolling volatility of the
different P\&Ls is shown in the left panel of
Fig.~\ref{rollvol}. Fluctuations around the average (here set to 1)
are quite large for all implementations, but the Markowitz one
appears to be better controlled. Some more insight on the risk control
can be deduced from the rolling correlation with a market-cap weighted
index, as shown in the right panel of Fig.~\ref{rollvol}. Here, again,
the FF approach results in very large fluctuations around zero, at
odds with the other portfolio construction techniques.

\begin{figure}[]
   \includegraphics[width= 0.45\textwidth]{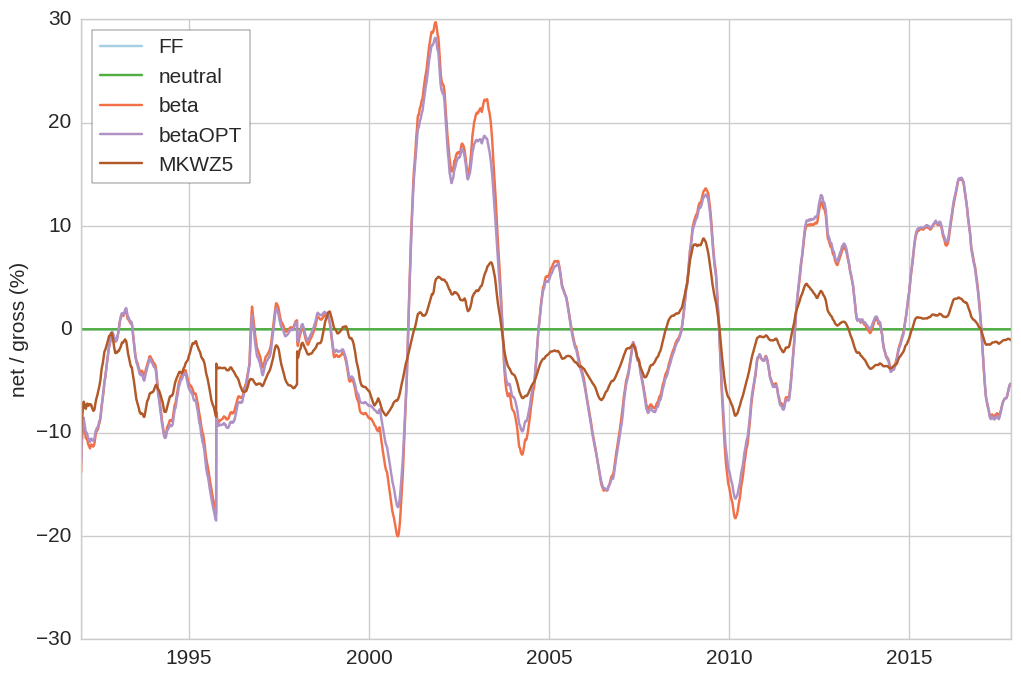}
   \includegraphics[width= 0.45\textwidth]{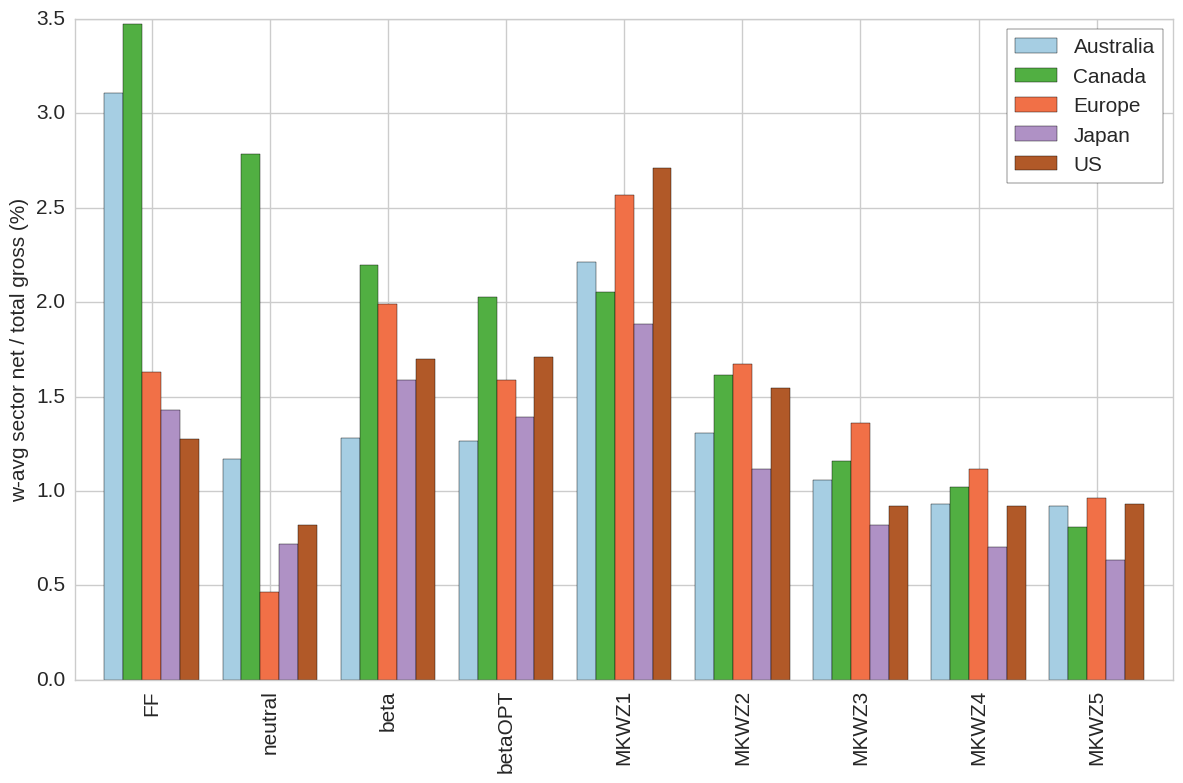}
   \caption{The average net market value exposure (left: global;
     right: sector) }
   \label{exposure}
\end{figure}

From a risk-control perspective, the global and sector exposures are
also interesting (see Fig.~\ref{exposure}). The net-over-gross market
value as a function of time is plotted in the left panel. The neutral solution has
its net equal to zero by construction, while all the other exhibit
fluctuations around zero. Once again, the fluctuations of the
Markowitz portfolio are smaller on average. The right panel focuses on
the average exposure on the classical industry sectors for the
different portfolios. The results are shown by pool. While FF does not
control at all these exposures, the neutral and Beta constructions
manage to reduce them to some extent. 

It is interesting to look at the
behavior of the Markowitz solution as a function of the number of
statistical factors kept in the optimization process (i.e. the
parameter $k$ in Eq.~\ref{clip}). The Markowitz $k=1$ solution, as discussed
above, is not very different from other $\beta$ neutralisation recipes. 
As we retain more statistical factors, we manage to better control
(albeit indirectly) the average sector exposure. This monotonous
behaviour is indeed observed in the plot. As sectors behave more like
indexes than like stocks, and thus exhibit a negative skewness, we
expect this empirical fact to have consequences on the tail
distribution of the P\&L as well (see next section).

\section{Momentum: Skewness and Tail Correlation}

The results shown so far suggest that the better performance of the
Markowitz approach is a consequence of a tighter control of the risk
of the portfolio, at least in terms of bulk metrics like variance,
correlations, and average exposures. 

But what about the skewness or the tail risk of the strategy? This is known to be an issue for many
equity factors, with Momentum as a case in point. More precisely, the Momentum tail
risk is mostly due to strong market rebounds. For these reasons,
we will now look explicitly at the skewness and the tail correlation of Momentum
with the market. 

\begin{figure}[]
   \includegraphics[width= 0.45\textwidth]{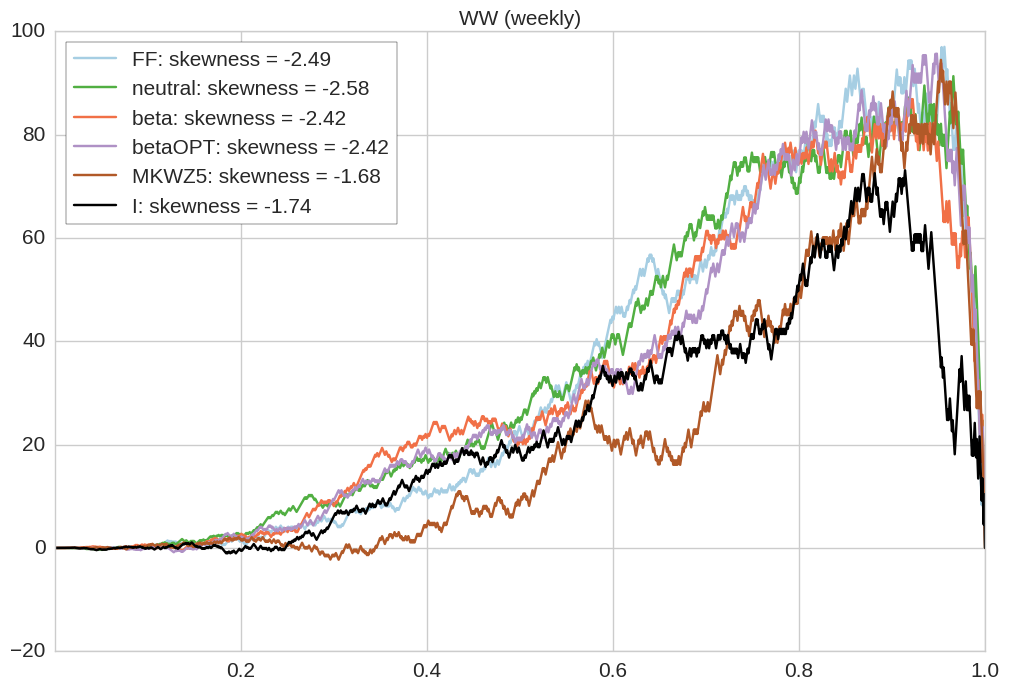}
   \includegraphics[width= 0.45\textwidth]{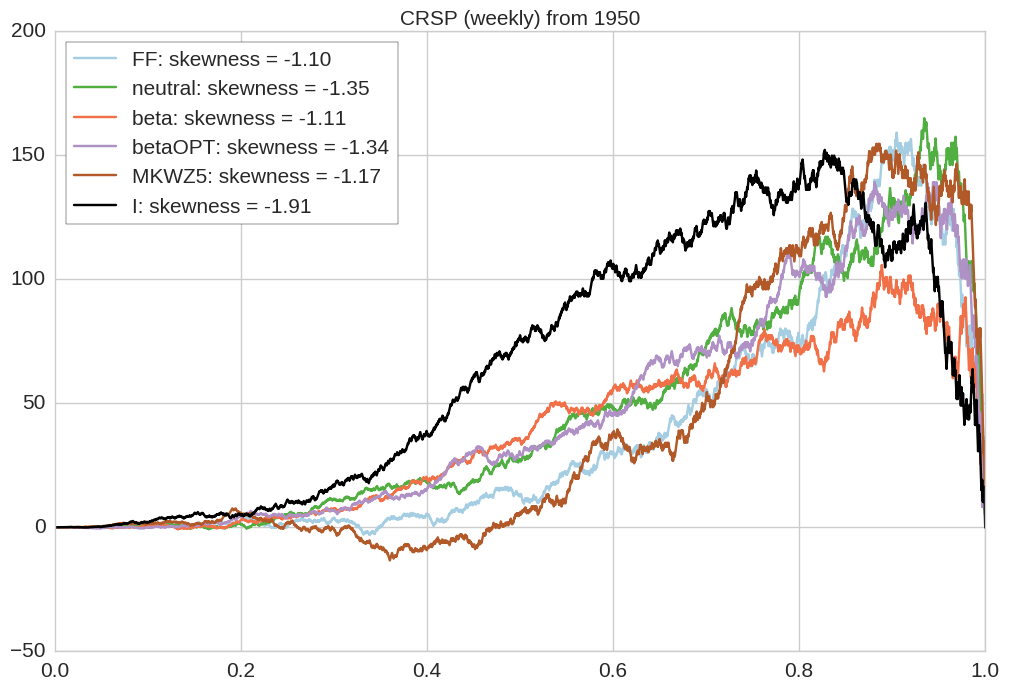}
   \caption{The weekly P\&Ls of the different strategies are reordered
     based on the amplitude, from the smallest to the largest in
     absolute value. The skewness can be computed directly from these
     data (\cite{Yves2017}) and is shown in the legend. Left: world-wide
     back-tests over the last 20 years. Right: US data since 1950. }
   \label{skewplot_CRSP}
\end{figure}

Fig. \ref{skewplot_CRSP} shows the weekly skewness of the world-wide
(left) and US long term (right) Momentum strategy. We follow the
approach introduced in~\cite{Yves2017} to visualize and compute the
skewness of a P\&L. The idea is to reorder the weekly P\&L data based
on the amplitude (i.e. absolute value) of the weekly P\&L itself. This
allows one to easily visualize how negative (if at all) are the
contributions of the most volatile weeks. Or, equivalently, what would
be the total P\&L if one was able to manage or hedge away the largest
contributions (positive or negative).  The market-cap index, which is
known to be strongly negatively skewed, is also shown for comparison
on the same graphs. We find a confirmation of the negative skewness of
the Momentum strategy; moreover, the cash- and $\beta$-neutral
approaches have heavier negative tails than FF and Markowitz.

One can learn more about the tails of the distribution by
investigating the strategy's behaviour when the market-cap weighted
index is either at a local minimum, or at a local maximum.  Momentum is known
to perform poorly when the market rebounds after a long-lasting
draw-down. The usual explanation relies on the role of Value investors
which step in after a long crisis and massively buy the stocks that
look cheap based on some fundamental metrics. These stocks probably
went down more than the average in the recent past and as such are
more likely to sit on the short leg of a Momentum portfolio. We will
now confirm that this picture is broadly correct, but once again it is
dependent on the portfolio construction.

\begin{figure}[]
   \includegraphics[width= 0.3\textwidth]{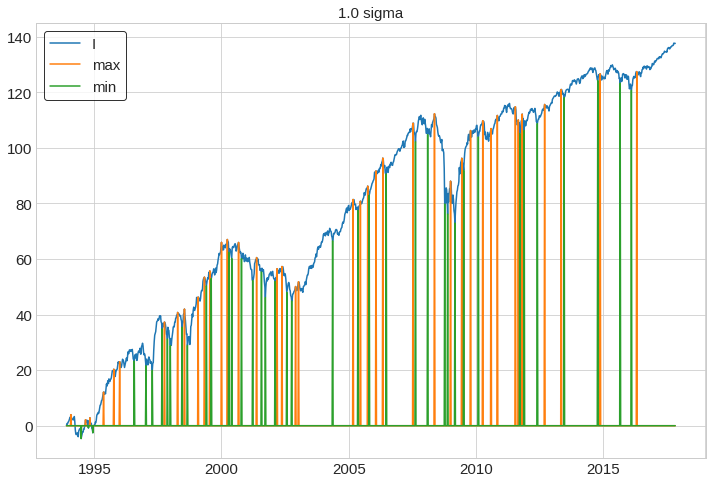}
   \includegraphics[width= 0.3\textwidth]{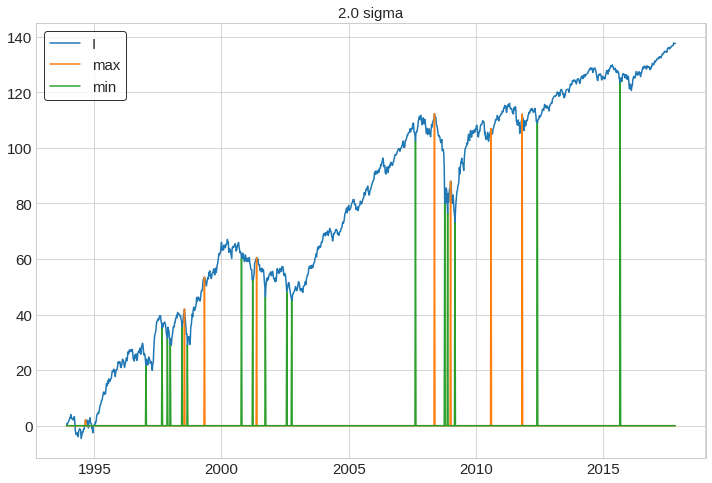}
   \includegraphics[width= 0.3\textwidth]{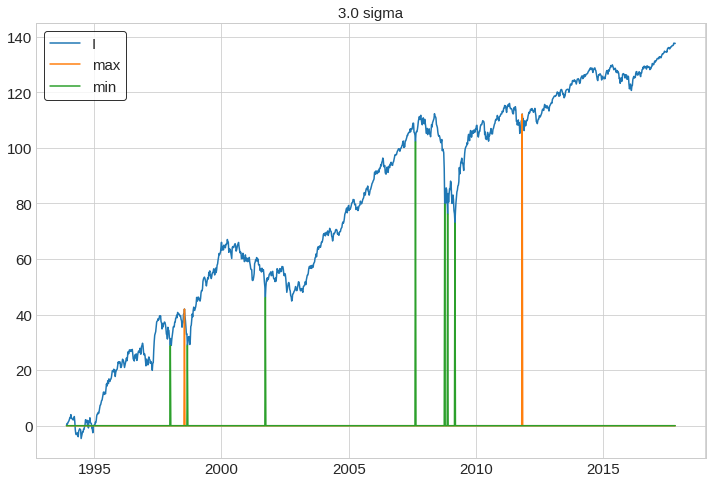}
   \caption{Operational definitions of market kink for different
     values of $\sigma$. All the plots show a World Wide market index
     with vertical bars placed in corresponence of local maxima/minima
     identified by the following procedure. We first find local minima 
     over periods of 9 weeks. If the corresponding draw-down is larger than 
     $n$ times the weekly volatility, then we retain the subsequent 4 weeks to
     compute the strategy's statistics in Fig.~\ref{pnl_kinks}. Local
     maxima are identified in a similar way. From left to right we show
     resulting kinks for respectively $n=1$, $n=2$ and $n=3$.}
   \label{kinks_sigma}
\end{figure}

To this end, we introduce an operational definition of market kinks,
which is illustrated in Fig.~\ref{kinks_sigma} and explained in the
corresponding caption. The relevant parameter here is the depth of the
local draw-down (or draw-up) measured in units of the (causal) rolling
volatility. We then compute, for different values of this parameter,
the average performance of the strategy over the next month. Here we
want to quantify the extent to which the negative skewness is related
to market rebounds, and how it depends on the portfolio construction.

\begin{figure}[]
   \includegraphics[width= 0.45\textwidth]{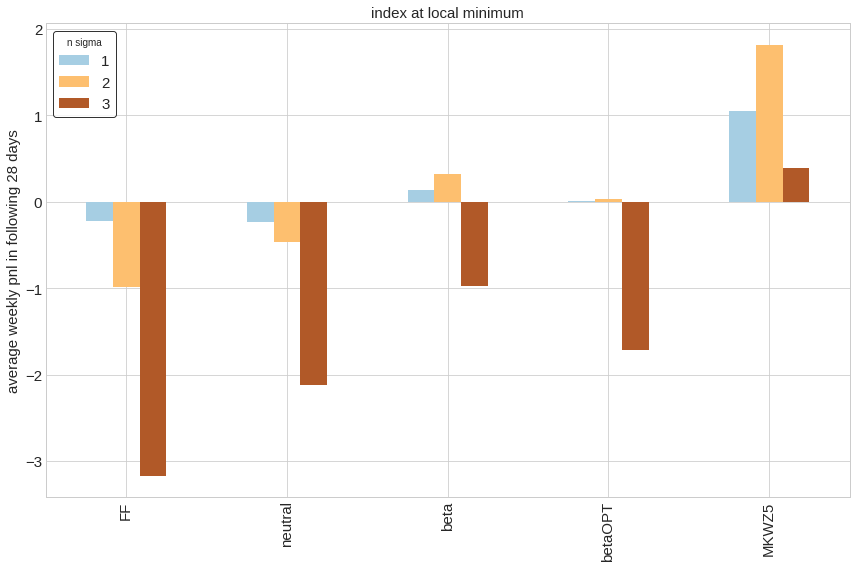}
   \includegraphics[width= 0.45\textwidth]{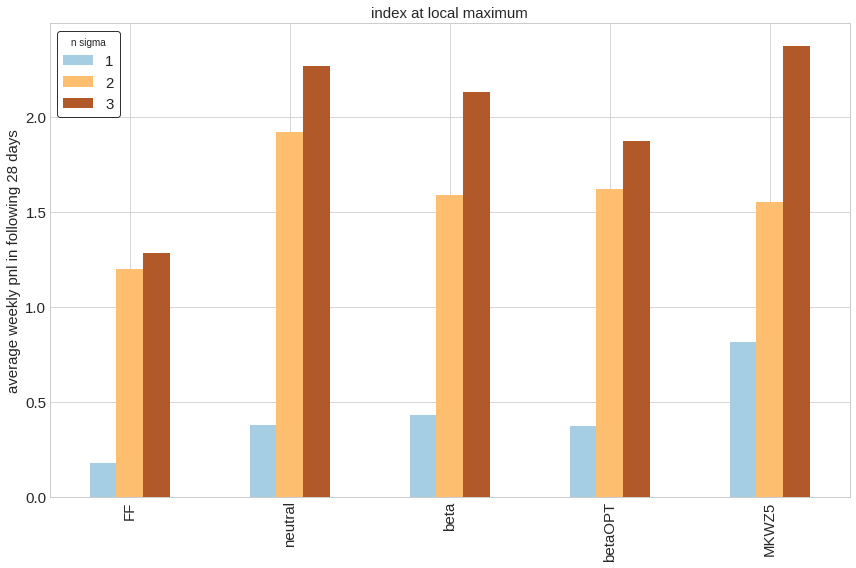}
   \caption{Left: The bars represent, for different values of the
     parameter $n$ described in the caption of Fig.~\ref{kinks_sigma},
     the average strategy's performance conditioned to a market
     rebound. See text for more comments. Right: The strategy's
     performance in periods of strong market corrections.}
   \label{pnl_kinks}
\end{figure}

In the left panel of Fig.~\ref{pnl_kinks} we see that all the
score-based portfolio schemes except the Markowitz portfolio
perform negatively after the index rebounds from a local minimum 
while the Markowitz portfolio is slightly positive.
This feature in
fact plays a major role in explaining why the overall performance is
better.  Moreover, the Markowitz portfolio is also among the best
performing portfolios after the market hits a local maximum (right
panel).


\section{Other Equity Factors}

The main message of the previous sections is that the Markowitz
portfolio construction is, for a number of reasons ranging from risk
control to tail correlation management, the best way to extract returns
from a Momentum signal. But what about the other equity factors? In
other words, how universal is this result?

\begin{figure}[]
   \includegraphics[width= 0.75\textwidth]{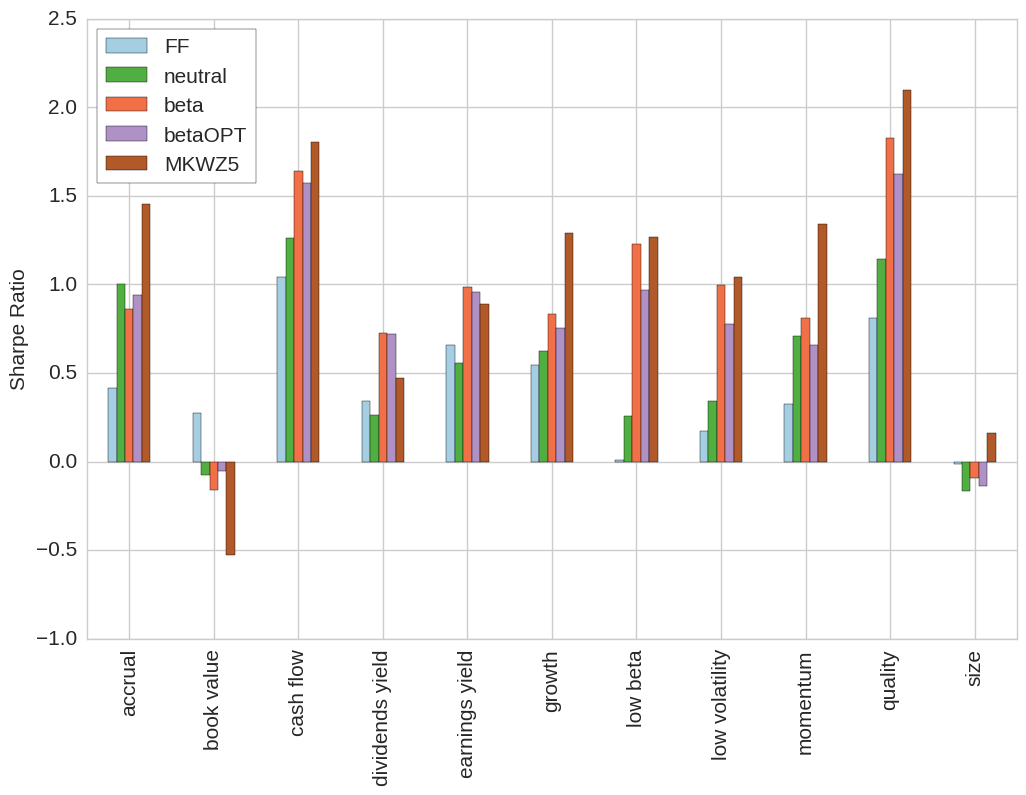}
   \caption{The world-wide Sharpe Ratio for the main equity factors,
     and for all the portfolio constructions.}
   \label{sharpe_all_factors}
\end{figure}

We have back-tested (using the same data and the same approach as
before) many other fundamental equity factors and the results are very
similar. We apply standard definitions of factors as can be found in
the literature:
\begin{itemize}
\item[Accrual:] is computed as the yearly increase of Net Operating
  Assets normalized by total assets ranked from lowest to highest.
\item[Book value:] total equity / market capitalization (lagged by 1
  month) and ranked from lowest to highest.
\item[Cash flow:] Operating Cash flow / market capitalization (lagged
  by 1 month) and ranked from lowest to highest.
\item[Dividend yield:] Dividends / market capitalization (lagged by 1
  month) and ranked from lowest to highest
\item[Earning yield:]  Net Income / market capitalization
  (lagged by 1 month) and ranked from lowest to highest
\item[Growth:] Cash Flow / total asset (lagged by 1 month) and ranked
  from lowest to highest.
\item[Quality:] Net Income / total assets (lagged by 1 month) and
  ranked from lowest to highest.
\item[Low Beta:] the beta of each stock is computed from the first statistical
  factor (i.e. the Beta vector is the $\sqrt{\lambda_0}\bm{v_0}$) and
  ranked from highest to lowest.
\item[Low Vol:] we compute a $180$ days rolling volatility
  (lagged by 1 month) and rank from highest to lowest.
\item[Momentum:] we compute $230$ days rolling average returns (lagged
  by 1 month) and rank from lowest to highest
\item[Size:] we compute $250$ days rolling averages market
  capitalizations (lagged by 1 month) and rank from highest to lowest
\end{itemize}
Fig.~\ref{sharpe_all_factors} shows that the Markowitz portfolio
construction generates the highest Sharpe Ratio 8 times out of 11, and
is neck-and-neck with Beta 2 times out of 11. The three ``bad'' cases
are Dividend yield and Earning Yield, and in particular Book-value,
whose performance is negative for all portfolio constructions. In some
good cases, like for Momentum or the Accrual anomaly, the effect is
quite spectacular. As for the Size effect, it seems to be the only way
to get a positive Sharpe Ratio from the small-cap premium, although
statistical significance may be debated.

Another interesting observation is that the low-risk models
(Low-Vol and Low-Beta) can be significantly improved by simply choosing
$\beta$-neutrality over cash-neutrality. This makes sense as long
positions on less risky stocks need to be leveraged in order to
achieve market neutrality, and that mechanically generates the net long
exposure needed by the model (see \cite{Beveratos17} for more details on
this point).

\section{Introducing transaction costs}

In this section we perform one more step towards reality as we include
transaction costs in the portfolio construction problem. In
the previous sections, this point was completely neglected and every
day the portfolio was supposed to be built from scratch with no memory
of previous positions. This is clearly unrealistic, so we briefly
touch on this point now.

When executing orders, one faces a number of different costs: we can
group them into three categories. There are commissions, including
broker and exchange fees; linear costs, related to the bid-offer
spread; and slippage costs, which are super-linear and take market
impact into account. In a simplified set-up, we will neglect slippage
costs as these would require a dedicated section on how we model them, and
how we calibrate the parameters in the model. We will take all the
linear costs into account: commissions, and half-spread costs.

\begin{figure}[]
   \includegraphics[width= 0.85\textwidth]{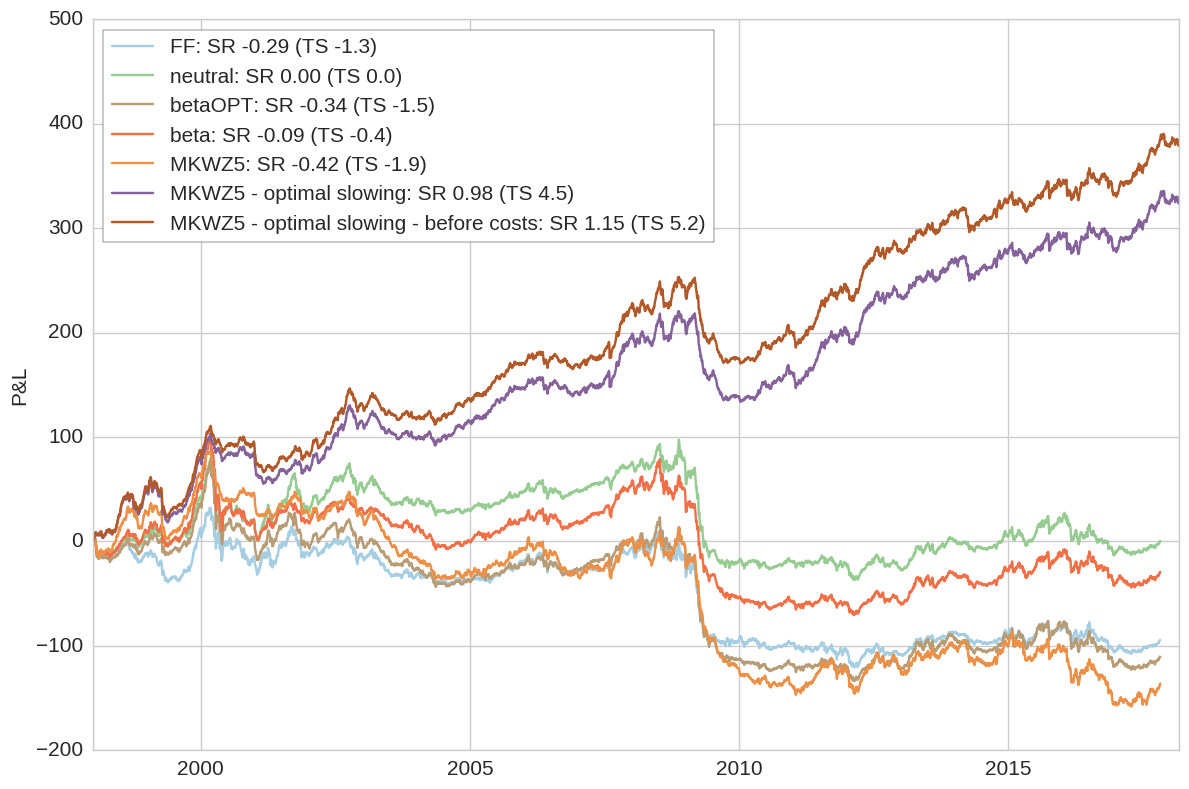}
   \caption{The first five curves represent the world-wide P\&L shown
     in Fig.~\ref{PnLWW} when we also include execution costs in the
     portfolio valuation. We observe a huge deterioration in the
     performance. The last two curves show the results (before and
     after costs) of a generalized mean-variance optimization problem
     where these executions costs are explicitely taken into
     account.}
   \label{moma_costs}
\end{figure}

When applied to the Momentum' portfolio mark-to-market, these costs
have a strong impact on the total performance (see
Fig.~\ref{moma_costs}). The Sharpe Ratio actually becomes negative for
all the approaches. This is due to the fact that these costs were not
considered by the algorithm and the daily turnover, as a consequence,
is artificially high. The key observation is that the Markowitz
approach, as opposed to the others, allows for a generalization
of the optimization problem which takes these costs into
account. The mathematical formulation of the problem gets more
involved and the details can be found for instance in~\cite{abeille16,pedersen13}. Here
we report the main results.

The last curve in Fig.~\ref{moma_costs} (brown colour) is the result
of the generalized Markowitz problem. In practice, the algorithm finds
the optimal frequency at which the predictor should be slowed-down
(similar to an exponential moving average) in order to maximise the gain
left after execution costs are removed. As a result, the Sharpe Ratio
computed before applying the cost formula is a bit worse than the
``pure'' one (1.15 vs 1.19). The main advantage is that if we now
include the costs in the mark-to-market, the Sharpe Ratio gets reduced
by only 15\% and we are left with SR=1.

\section{Conclusions}\label{conclusions}

In this paper we have shown that the portfolio construction process
plays a crucial role in the implementation of an equity market
neutral strategy. Although most of the details have only been shown for the
Momentum strategy, the main results hold true for many classic equity
strategies. We have shown that the portfolio construction step may
lead to quite large differences in the resulting performance. We
believe that this explains to a large extent the dispersion usually
observed among market neutral managers. 

As discussed in the paper, the Markowitz approach, once the
correlation cleaning problem is properly dealt with, is the best way
to deal with complex covariance structure in the single-name security
space. This is clearly observed from a performance and risk
perspective, whether bulk or tail metrics are considered.

Of note, the Markowitz approach lends itself to a straightforward
generalization to the multi-signal case, at odds with e.g. the FF
scheme which involves creating multiple embedded buckets of stocks
where the ordering of signals plays a role in the construction of the
portfolio.

We have also addressed the question of transaction costs, and how to
build a portfolio construction algorithm that would slow-down the
predictor in order to make the system trade as little as possible. It
turns out that the main results of the present paper generalize when
one uses a dynamical Markowitz approach that accounts for
costs~\cite{abeille16, pedersen13}. 

\emph{Acknowledgements.} The authors would like to thank Jean-Philippe
Bouchaud, Yves Lemp\'eri\`ere, Philip Seager, Emmanuel S\'eri\'e, Matthieu
Cristelli, Guillaume Simon, Charles-Albert Lehalle and Andr\'e Breedt for insightful 
comments and fruitful discussions.

\bibliography{biblio}{}
\bibliographystyle{plain}

\end{document}